\documentclass[namedreferences]{SolarPhysics}
\usepackage[optionalrh]{spr-sola-addons} 
\usepackage{graphicx}                    
\usepackage{color}                       
\usepackage{url}                         
\begin{document}                                
\begin{article}
\begin{opening}         
\title{Enhanced {\it p}-Mode Absorption seen near the Sunspot 
Umbral-Penumbral Boundary}

\author{Shibu K. Mathew}
\runningauthor{S.K.Mathew}
\runningtitle{Enhanced p-Mode Absorption near Umbral-Penumbral Boundary}

\institute{Udaipur Solar Observatory, Physical Research Laboratory, Badi Road, Udaipur - 313001, India\\
\email{shibu@prl.res.in}}

\begin{abstract}
We investigate {\em p}-mode absorption in a sunspot using 
SOHO/MDI high-resolution Doppler images. The Doppler power 
computed  from a three and a half hour data set is used 
for studying the absorption in a sunspot. The result 
shows an enhancement in absorption near the umbral-penumbral 
boundary of the sunspot. We attempt to relate the observed 
absorption with the magnetic field structure of the sunspot.      
The transverse component of the potential field is computed
using the observed  SOHO/MDI line-of-sight magnetograms.  
A comparison of the power map  and the computed potential 
field shows enhanced absorption near the umbral-penumbral 
boundary where the computed transverse field 
strength is higher.    
\end{abstract}      
\date{Received ......................; accepted ......................}

\end{opening}
\section{Introduction}
The interaction of solar oscillations with magnetic field has been
reported by many authors. Two of the important findings are the
reduced {\em p}-mode power in regions of strong magnetic field
and the enhancement of power in higher frequencies surrounding
the regions of strong fields (Woods and Cram, \citeyear{woods};
Tarbell {\it et al}., \citeyear{tarbel}; Brown {\it et al}., 
\citeyear{brown}). The  magnetic field reduces {\em p}-mode power 
in active regions in the 3 mHz band while enhances power of 
these modes in the 5 mHz band (Hindman and Brown, \citeyear{hindman}; 
Venkatakrishnan, Kumar, and Tripathy, \citeyear{pvk}).   
The mechanisms of {\em p}-mode absorption were reviewed 
by Spruit (\citeyear{spruit}) and he suggested a promising 
mechanism for reduced power as the conversion of {\em p} mode 
into a downward propagating slow mode along the magnetic flux 
tubes (Spruit and Bogdan, \citeyear{spruit1};
Cally and Bogdan, \citeyear{cally1}; Cally, Bogdan, 
and Zweibel, \citeyear{cally2}). The waves in the flux tube, 
once excited by the sound wave, can carry energy
out of the {\em p} modes through the  wave guide into the convection 
zone, thereby  producing a reduction in observed power. Simulations by 
Cally (\citeyear{cally}) show that the enhanced absorption  takes place 
primarily in the more inclined magnetic field regions towards the edge of 
the spot. A comparison of the spatial distribution of Doppler power 
and the magnetic-field configuration in a sunspot could reveal 
the variation of Doppler power with magnetic-field strength 
and the field inclination. In this paper, we study {\em p}-mode
absorption in magnetic field concentrations and compare that with 
the longitudinal and the computed transverse field configuration in a sunspot.

\section{Data Sets}
For this analysis, we have used high-resolution $(0.6''/{\rm pixel})$
Dopplergrams with one-minute cadence obtained with the {\it Solar and
Heliospheric Observatory} / Michelson Doppler Imager (SOHO/MDI) 
instrument (Scherrer {\it et al}., \citeyear{scherrer}). Complemented 
with these data, we have also used few line-of-sight magnetograms and 
intensity images from SOHO/MDI. The  sunspot analyzed is a member of 
the active region NOAA 8395 and the observations were made on 1 December 
1998 between 04:00 UT and 07:51 UT. The  sunspot was near  disk center 
($\mu \approx 0.93$). Images were registered and re-mapped onto 
heliographic coordinates. Registration is carried out first in  
synodic rate and then by cross-correlating  successive images 
with a reference image. The reference image is updated after 
every five minute to avoid errors in registration resulting 
from the evolution of  the active region. The resulting images consist 
of $773 \times 393 $ pixels corresponding to $\approx 8 \times 4$ 
arc-min at the solar disk center. We selected a small sub-region
of $150 \times 100$ including one of the sunspots in the active 
region for our analysis. 
\begin{figure}
\centerline{\includegraphics[width=30pc]{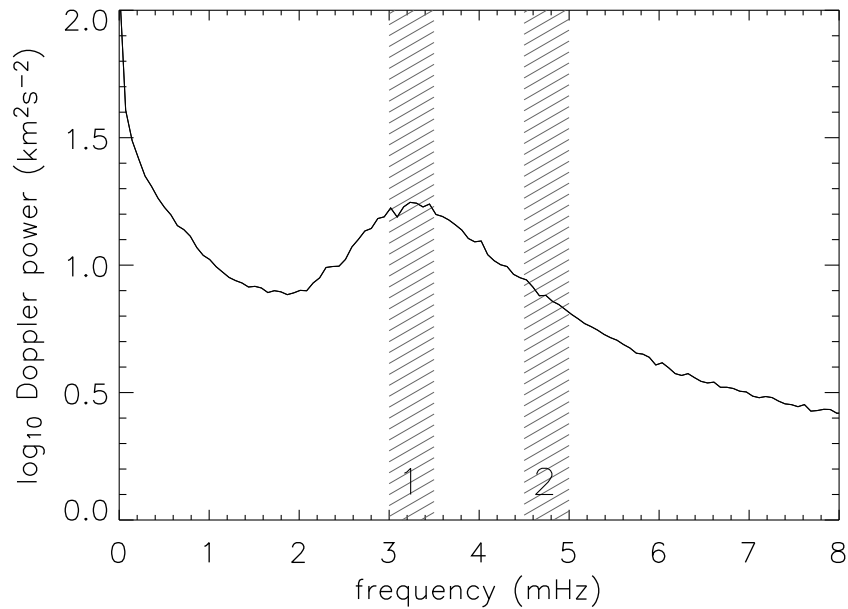}} 
\caption{The spatially averaged Doppler power spectrum for the observed 
region. The hatched areas correspond to the frequency bands between 
3\,--\,3.5 mHz (band 1) and 4.5\,--\,5 mHz (band 2).}
\end{figure}

The duration of 3:51 and the cadence of one minute give a frequency 
resolution and Nyquist frequency of 72.1 $\mu {\rm Hz}$ and 8.33 mHz, 
respectively. Re-mapped SOHO/MDI Dopplergrams and line-of-sight 
magnetograms were used for comparing the {\em p}-mode absorption 
with magnetic-field strength. An average of two magnetograms, one 
at the beginning and other towards the end of the observing period 
are used for the comparison. 

\section{Data Analysis}
We used a method similar to that described by Brown {\it et al}. 
(\citeyear{brown}) for the Dopplergram analysis. 
From the accurately registered Doppler images, 
a time series is constructed. The power spectra is computed for 
every pixel in the image.  Figure 1 shows the amplitude of the 
average power spectrum for the above mentioned sub-region including 
the sunspot. The average power within the two frequency bands represented 
by the hatched areas are used for obtaining the results shown in the 
subsequent figures. The range of values included in the low (band 1) 
and high (band 2) frequency bands are 3\,--\,3.5 mHz and  4.5\,--\,5 mHz, 
respectively.

From the line-of-sight magnetograms, the potential-field configuration 
of the entire active region is computed. All three components of 
the magnetic field ($B_{x}, B_{y}$ and $B_{z}$) are derived assuming 
the potential-field approximation using the Fourier method (Sakurai, 
\citeyear{sakurai}). The mean field has been removed before 
computing the magnetic-field components  and later added.
The normal component ($B_{z}$) of the magnetic-field is needed as 
the boundary condition for the potential-field extrapolation, whereas 
only the line-of-sight field is available from the MDI measurements. 
The measured line-of-sight field and the normal component 
differ when the sunspot is observed away from the disk center due 
to the projection effect. In our analysis a first order correction 
is carried out on the line-of-sight magnetic field 
for the projection effect taking the heliocentric angle into 
consideration ({\it i.e.} $B_{z} = B_{LOS}/\cos \theta$). 
Since the observation of the analyzed sunspot were obtained 
when it was  close to the disk center $(\mu = \cos \theta = 0.93)$,  
the above correction is approximately valid for strong normal 
magnetic-field component. The transverse magnetic-field component 
is $\sqrt{B_{x}^2+B_{y}^2}$ and the magnetic-field inclination is
$ \tan^{-1}(B_{z}/\sqrt{B_{x}^2+B_{y}^2}$). Even though the computed 
potential field assumes the lowest magnetic-energy configuration of 
the sunspot, for obtaining a rather simplified idea about the magnetic 
structure the potential field calculation can be used. The computed 
magnetic-field structure (Figure 2c) of the observed sunspot is similar 
to the retrieved transverse field configuration of sunspots from the 
vector magnetic-field measurements (Westendorp Plaza {\it et al}., 
\citeyear{westen}; Keppens and Martinez Pillet, \citeyear{keppens}; 
Mathew {\it et al}., \citeyear{mathew}).  
\begin{figure}
\centerline{\includegraphics[width=26pc]{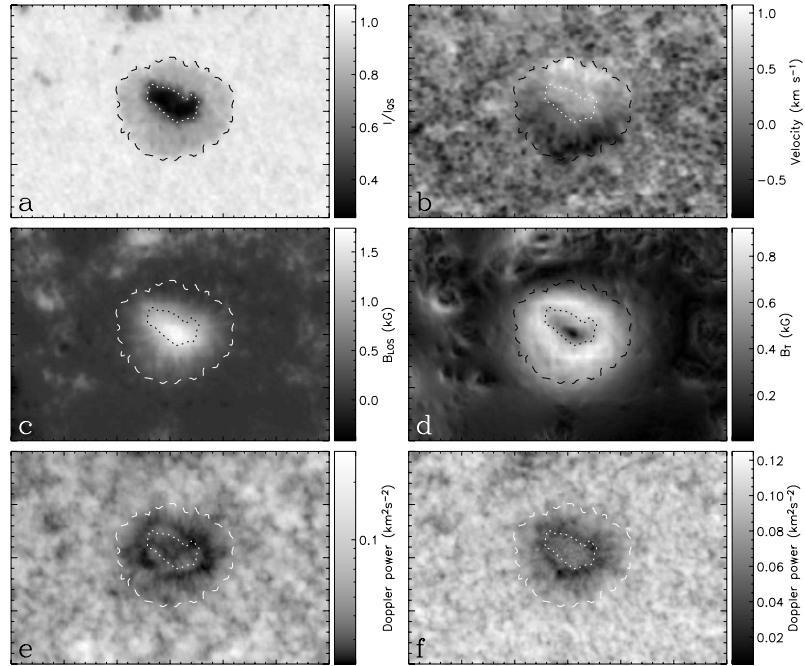}}
\vskip 0.2cm 
\caption{Continuum intensity (a), line-of-sight velocity (b), line-of-sight magnetic field (c), 
computed transverse magnetic field (d), and the total Doppler power in the 1 (e) 
and 2 (f) frequency  bands for the analyzed sunspot. The dotted and dashed contours are the 
umbral and penumbral boundaries, respectively}
\end{figure}

\section{Results}
Figure 2 show the continuum image (a), Dopplergram (b), the line-of-sight 
magnetogram (c), and the computed transverse field (d) for the analyzed sunspot. 
Figure 2 (e) and (f) show the average Doppler power for the frequency bands 
1  and 2 marked in Figure 1. The Doppler power is displayed in 
logarithmic scale to make the absorption clear in the lower levels. 
The dotted- and dashed-line contours are plotted for the umbral and penumbral 
boundaries, respectively. The Doppler power in both lower (band 1) and  the 
upper (band 2) frequency bands (band 1) show an enhanced absorption near the 
umbral-penumbral boundary; here the computed 
transverse field also becomes strong. 
In the upper panel of Figure 3, the continuum image of the analyzed sunspot along 
with the equal intensity contours are displayed. The inset in the upper panel 
shows the brightness enhanced image of the sunspot. This gives a closer 
view of the splitted and irregular structure of the umbra. In the lower 
panel, variation of different parameters along two radial cuts  
(marked as {\it a} and {\it b} 
in the continuum image) through the sunspot are plotted. The dotted vertical 
lines in these figures mark the umbral boundary and the plotted Doppler power 
are for the lower-frequency band. Around the umbral boundary the line-of-sight 
magnetic filed shows a smooth change, while the computed transverse field strength 
reaches a maximum value. The Doppler power in the lower frequency band shows a dip 
around this location in both the radial cuts. 
\begin{figure*}
\centerline{\includegraphics[width=20pc]{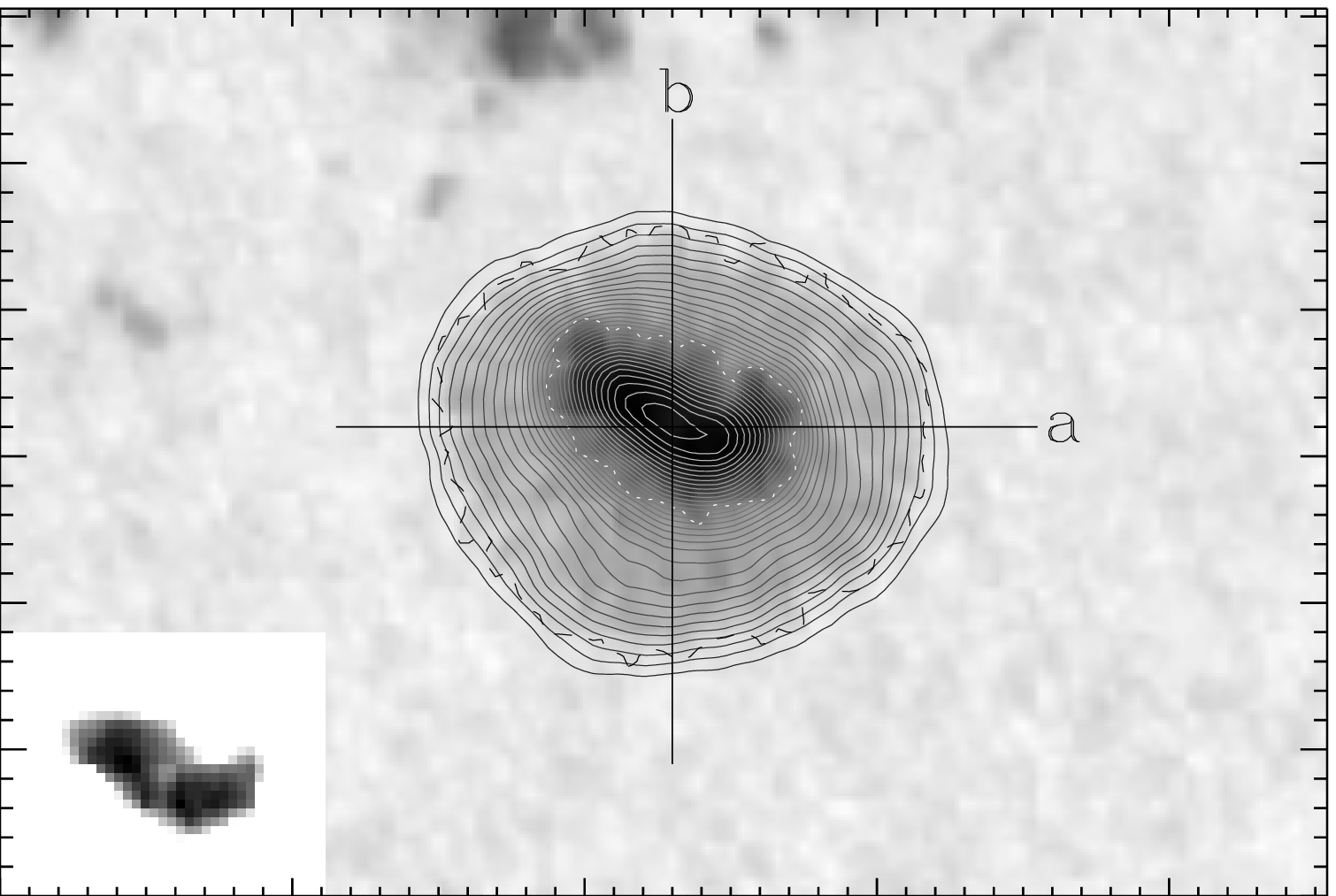}} 
\vskip 0.3in
\centerline{\includegraphics[width=28pc]{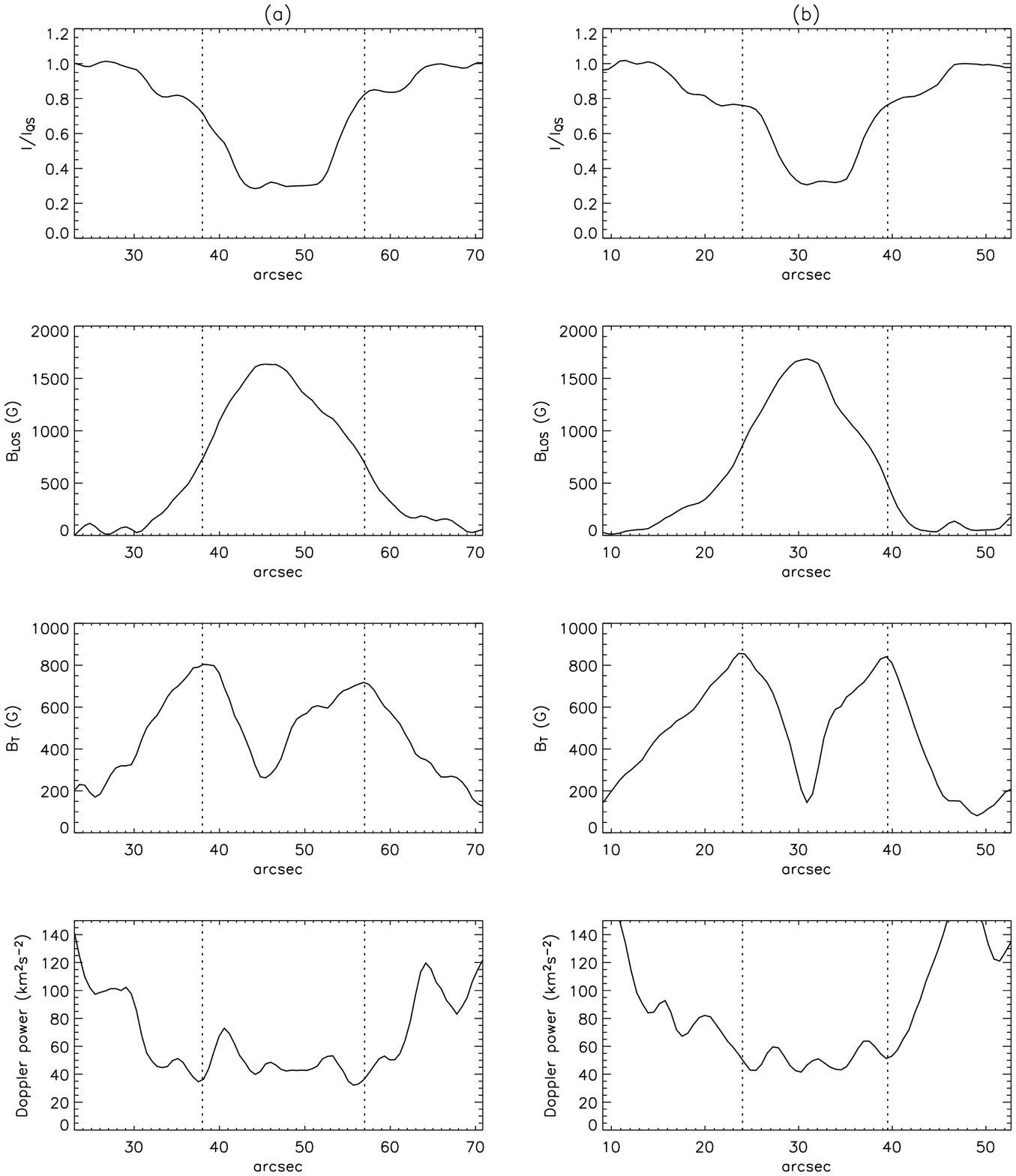}} 
\caption{Various parameters (lower panel) for two cuts across the sunspot indicated 
by straight lines {\it a} and {\it b} in the upper panel. The two vertical dotted 
lines show the umbral-penumbral boundary.  
The average values  between the adjacent solid contours are used to
obtain the radial distribution of these parameters, shown in Figure 4.}   
\end{figure*}

In Figure 4 we plot the values averaged between equal intensity contours for the 
continuum (a),  line-of-sight magnetic field strength (b), 
the computed transverse magnetic field (c), the computed inclination (d), 
Doppler power in the lower frequency band (e), and the Doppler power in 
higher frequency band (f). The contour numbers plotted are from the upper panel in Figure 3, 
starting from the inner umbra. Here also, a clear reduction in Doppler power near the umbral 
boundary is observed, where the strength of the computed transverse field is higher.      
\begin{figure}
\centerline{\includegraphics[width=28pc]{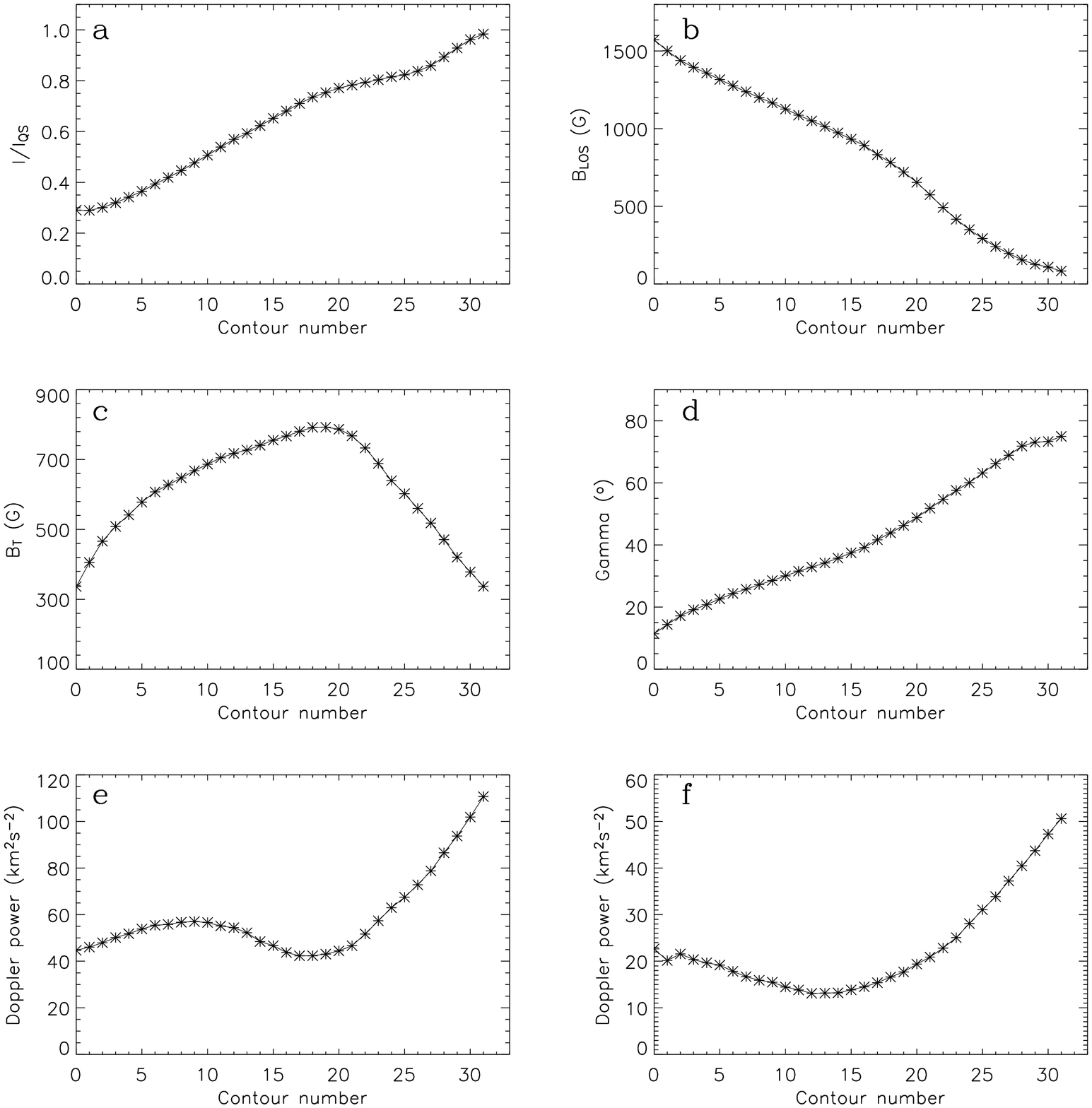}} 
\caption{Azimuthal averages of the intensity (a), line-of-sight magnetic field (b), 
computed transverse field (c), computed field inclination (d), Doppler power in 
band 1 (e), and Doppler power in band 2 (f)} 
\end{figure}

\section{Discussion}
It is a well known that the Doppler power is considerably reduced in  
sunspots compared with the quiet Sun. Braun, Duvall, and LaBonte (\citeyear{braun1}) 
reported that sunspots absorb up to half of the incident 
{\em p}-mode power at favored frequencies and horizontal wavenumbers. From the 
analysis of  the acoustic properties of two large sunspots using Fourier-Hankel 
decomposition of {\em p}-mode amplitudes, Braun (\citeyear{braun2}) showed that 
there is a peak in Doppler power absorption centered at around 3 mHz and an absence 
of absorption at 5 mHz frequencies. Lindsey and Braun (\citeyear{lindsey}) 
reported a ``penumbral acoustic anomaly" in which they found a ring of relatively 
depressed acoustic power in and around the sunspot penumbra. Study of active region 
oscillations by Muglach, Hofmann, and Staude (\citeyear{muglach}) also show (in their Figure 5) a 
reduction in Doppler power in the penumbra. 

In our analysis, we present the spatial distribution of the oscillatory power in a sunspot.
The total-power maps obtained for this sunspot in two frequency bands 
(3\,--\,3.5 mHz and 4.5\,--\,5 mHz) show a structured ring-like depression in the penumbra. 
The reduction is visible in both the frequency bands with a difference of a shift 
in the peak absorption towards the umbral-penumbral boundary in the higher frequency 
band. We also observe an absorption structure in the umbra in the lower frequency band 
which is not clearly visible in the higher frequency band. A comparison of the reduction in {\em p}-mode
power with magnetic structure of the sunspot clearly indicates increased absorption 
near the locations where the transverse field is higher. The absorption structure 
in the umbra (also the decrease in Doppler power in umbra seen in the azimuthal average, Figure 4), 
without correspondence with transverse-field structure could be due to the split structure of 
the umbra which is clearly seen in the brightness-enhanced image of the sunspot 
displayed in the upper panel of Figure 3. We suggest that this absorption structure 
could be due to the increased field inclination resulting from the split structure 
of the umbra, which is not visible in the MDI line-of-sight magnetogram and thus 
also not in the computed transverse field. 

Theoretical modeling of the influence of sunspot magnetic field on the incident 
oscillation has been described by several authors (Spruit, \citeyear{spruit0}; 
Spruit and Bogdan, \citeyear{spruit1}; Cally and Bogdan \citeyear{cally1}; 
Cally, \citeyear{cally}; Rosenthal and Julien, \citeyear{rosen}). 
In the presence of strong magnetic 
field, it is suggested that the fast wave is partially converted into 
a slow magneto-acoustic wave and is directionally influenced by the 
magnetic field. Simulations by Cally (\citeyear{cally}) showed
that the interaction of acoustic wave and magnetic field  has a 
strong dependence on the inclination of the magnetic field and found 
that the absorption peaks at 30$^{\circ}$ inclination angle 
(Cally, Crouch, and Braun, \citeyear{cally3}). In our analysis we find that 
the computed inclination angles are around 45$^{\circ}$ and 35$^{\circ}$,
where the absorption peaks in the lower (band 1) and higher (band 2) frequency
bands, respectively. These values should be considered cautiously since 
the exact inclination angle can only be derived from full 
vector magnetic-field measurements corrected for the projection effect. In our case the 
potential field and thus the field inclination is computed from the measured 
line-of-sight magnetic field and a simple approach is used for correcting 
the observed field before computing the potential components, assuming 
the magnetic field is perpendicular to the photosphere. 
Here, without having  full information on the vector magnetic 
field, the contribution of the transverse component to the line-of-sight field is unknown. 
But, it has been shown  that the difference between the computed 
potential field from the observed and the corrected (for the projection effect using full 
vector magnetic-field measurements) line-of-sight field is not very different from each 
other when the sunspot is located close to the disk center (Hagyard, \citeyear{hagyard}). 
From the analysis of an active region which is positioned within one-third solar 
radius from the disk center, Hagyard showed that the difference between the computed transverse 
field from the corrected and observed line-of-sight field is around 200 G and this difference 
occurs only in localized area of the umbrae of large sunspot. In our case, since the sunspot 
is close to the disk center, we presume that the simple correction which we employed is 
sufficient to represent the potential configuration of the sunspot.

More strikingly, we find that the computed 
transverse field reaches maximum values where the peak in absorption is observed. 
It is important to have full spectro-polarimetric vector magnetic-field observations 
to obtain a clear idea about the dependence of the field configuration on the spatial 
distribution of acoustic power. We have not carried out the study 
of effects of  magnetic field in the Doppler measurement which is 
done on the Zeeman sensitive spectral line profile. 
Rajaguru {\it et al}. (\citeyear{rajaguru}) presented  a detailed study of the phase 
shifts on the acoustic waves when observed in a sunspot. They have shown that 
the phases of the acoustic waves within a sunspot, especially in the penumbra, 
could be altered due to the propagating nature of the {\em p} modes and also 
due to the Zeeman split of the spectral line which could in effect produce reduction 
in Doppler power in a particular frequency band.             
                                            
\section{Conclusions}
We analyzed the spatial distribution of Doppler power in a sunspot observed near 
disk center. We find a structured ring like absorption 
pattern in Doppler power  near the umbral-penumbral boundary. The computed 
transverse field is higher at those locations where the peak depression in 
Doppler power is observed. The computed inclination angle ranges between 
35$^{\circ}$ and 45$^{\circ}$ at these locations. In order to understand the 
exact dependence of magnetic field strength and inclination on the Doppler 
power absorption, full vector magnetic-field observations are required. Also, 
it is preferred to have Doppler observation in magnetically insensitive lines 
to avoid any cross-talk between the line-of-sight velocity measurement and 
Zeeman splitting of the spectral line.  

\begin{acks}
I would like to thank B. Ravindra for providing the code for potential field calculation 
and the anonymous referee for their valuable comments which helped to improve the paper substantially.
SOHO is a project of international cooperation between ESA and NASA.
\end{acks}

\end{article}
\end{document}